\documentstyle[aps,prl,twocolumn,psfig]{revtex}
\begin{document}
\title{Inherent-Structure Dynamics and Diffusion in Liquids}
\author{T. Keyes and J. Chowdhary}
\address{{\it Department of Chemistry, Boston University, Boston, MA 02215}}
\date{(\today)}
\maketitle

\begin{abstract}
The self-diffusion constant $D$ is expressed in terms of transitions among the local minima of the potential ({\it inherent structures}, IS) and their correlations. The formulae are evaluated and tested against simulation in the supercooled, unit-density Lennard-Jones liquid. The approximation of uncorrelated IS-transition (IST) vectors, $D_{0}$, greatly exceeds $D$ in the upper temperature range, but merges with simulation at reduced $T \sim 0.50$. Since uncorrelated IST are associated with a hopping mechanism, the condition $D \sim D_{0}$ provides a new way to identify the crossover to hopping. The results suggest that theories of diffusion in deeply supercooled liquids may be based on weakly correlated IST.
\end{abstract}

\section{Introduction \label{sec-intro}}
The configuration ${\bf r}(t)$ of a liquid may \cite{sw} be mapped to the local minimum ${\bf R}(t)$, of the potential energy $U({\bf r})$, to which it drains; both vectors denote the 3N coordinates of a system with N atoms. Stillinger named the minima {\it inherent structures} (IS) and Stillinger et. al. have \cite{fssci} pioneered the application of this mapping to the theory of liquids. The configuration space decomposes into the basins of attraction of the IS, and so too does the configuration integral, providing a new approach to equilibrium statistical mechanics. Turning to dynamics, Stillinger and Weber \cite{sw} calculated the average T-dependent rate $<\omega_{is}(T)>$ of {\it IS-transitions} (IST) for an atomic liquid, obtained an activation energy comparable to that for diffusion, and generally discussed the connection between IST and physical dynamics.

With the system described by both the continuously varying ${\bf r}(t)$ and the discontinuous ${\bf R}(t)$ an explicit realization of \cite{gold} Goldstein's ideas about dynamics over the potential energy {\it landscape} is obtained. The most challenging area of liquid theory is the supercooled state, where the landscape paradigm is widely accepted. Below a critical temperature $T_{c}$ the system is believed to remain in a basin for relatively long times, with infrequent activated hops to neighbor basins. At $T>T_{c}$ motion is freer and not activated. The character of dynamics has been related to \cite{sri} the $T$-dependence of the IS energy, and to \cite{fssci} the roughness of the landscape. Nevertheless, no detailed calculation the self-diffusion constant D has been given. In this Letter we demonstrate the relation between D and IST in supercooled unit-density Lennard-Jones (LJ). At the lowest attainable $T$, $D$ is described accurately by a simple Markov approximation, and a qualitative change in the dynamics is evident.

\section{Diffusion and Inherent Structure Dynamics \label{difis}}
The diffusion constant is determined by the long-time linear behavior of the mean-square displacement (MSD). Since the magnitude of the \cite{lav} {\it return distance} ${\bf u}(t) = {\bf r}(t) - {\bf R}(t)$ is bounded by the size of a single basin while the MSD increases without limit the IS-MSD also serves, $<(\Delta {\bf R}(t))^{2}>/6N = Dt, t \rightarrow \infty$, where  $\Delta {\bf R}(t) = {\bf R}(t) - {\bf R}(0)$. The IS-MSD is the sum of the IST vectors, the separations of successive IS, $\Delta {\bf R}(t) = \sum_{\alpha=1}^{n(t)}{\bf \delta R}_{\alpha}$ after $n$ transitions. Squaring, averaging for fixed $n$, averaging over the distribution of $n$ at time $t$ and dividing by $6N$ yields
\begin{equation}
D = [<({\bf \delta R}^{2}) \omega_{is}> + 2 \sum_{\beta=1}^{\infty} <({\bf \delta R_{\alpha} \cdot \delta R}_{\alpha+\beta}) \omega_{is}>]/6N, \label{ismsd}
\end{equation}
where the sum is the correlation function $C(\beta)$ for $\beta$-neighbor IST vectors and we have assmed that $t$ is much longer than the persistence time of any correlations, required for diffusive behavior. Approximations to $D$ keeping $\beta \leq m$ only are denoted $D_{m}$ and the IS-Markov approximation is $D_{0}$, the first term on the RHS. The IST vector correlation $C(\beta)$ is somewhat analogous to the velocity correlation.

With large times and numbers of trajectories the averages in Eq~\ref{ismsd} factorize into $<\omega_{is}>$ times averages of ${\bf \delta R}$ only, and in particular the Markov approximation yields $D_{0} = <{\bf \delta R}^{2}> <\omega_{is}>/6N$. In computer simulation of supercooled liquids fluctuations are large and we use Eq~\ref{ismsd} as is in averaging over multiple MD runs; nevertheless the difference is within the noise and our results may be understood via the factorization approximation. To see how $D$ achieves its status as an intensive quantity consider that a liquid with short-ranged correlations may be \cite{ag} roughly divided into independent local regions. Successive IS differ \cite{sw} in the coordinates of a small number of particles and $<{\bf \delta R}^{2}> \sim O(1)$; IST are local rearrangements. On the other hand the number of regions is $O(N)$ and a change in any region changes the IS of the entire system, so \cite{sw} $<\omega_{is}> \sim O(N)$ and $D \sim O(1)$. The N-dependence of $<\omega_{is}>$ indicates that IS-dynamics studies should use the smallest realistic system to avoid an intractable IST rate with no direct correspondence to a physical rate.
\section{Simulation Methods and Results\label{sim}}
 We have previously studied \cite{tk} the unit-density supercooled LJ liquid, and use it with $N=32$, and with methods from prior work, to test the relation between IS-dynamics and diffusion. Natural LJ units are used throughout, well depth $\epsilon$ for T, hard core radius $\sigma$ for distance, time unit $\tau_{LJ} = (m\sigma^{2}/\epsilon)^{1/2}$ (2.18 ps for argon) and $m$ is the mass; the crystal melts at $T \sim 1.6$. A hot liquid at $T=5.00$ is cooled in one step to a temperature in the 1.20-0.60 range. The system is equilibrated for 2.5 $\tau_{LJ}$, data are gathered for 62.5 $\tau_{LJ}$, $T$ is decreased by 0.02 and the process is repeated 10-25 times, generating a single quench run. The cooling rate is $3.8X10^{-4}$. Different $T=5.00$ configurations lead to different behavior, and thus we obtain an ensemble of quenches. At $N=32$ the abrupt drops in $U$ signaling solidification, common at $N=256$, do not occur but some quenches develop solid-like pair distributions and these are discarded. Results are averaged over all surviving quenches, typically 5-15, at each $T$. Quench-to-quench fluctuations are much larger than any systematic changes over $T=0.02$, so we also average results at each $T$ with those from the next higher and lower $T$. Even so our data are somewhat noisy but the trends are clear for $1.10>T>0.34$. For now $T=0.34$ is the lower limit  because of the scarcity of liquid-like states and the magnitude of fluctuations compared to the mean in dynamical quantities ($D, \omega_{is}$).

\begin{figure} 
\psfig{figure=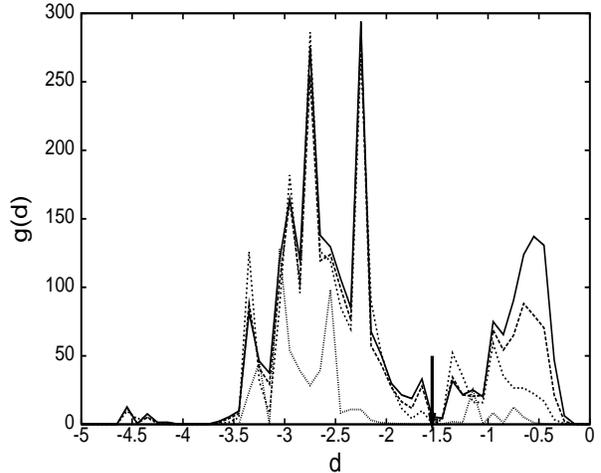,height=2.5in,width=3in,angle=-90}
\caption{Distributions of logarithms of IS-transition distances. Top to bottom at $d=-1,T=0.90, 0.60, 0.50, 0.40.$  \label{gofd}}
\end{figure}

Conjugate gradient (CG) minimizations are performed every 5 time steps  ($dt=.00125$), or 160 minimizations/$\tau_{LJ}$. Since the range of $<\omega_{is}(T)>$ is from 9.6 IST/$\tau_{LJ}$ at T=1.10 to 0.28 at T=0.38, this should be sufficient. The determination of whether a transition has occurred is an important, nontrivial matter. We begin, following \cite{donati}  Donati et. al., by calculating the distribution $g(d)$ of distances $d=log[(({\bf R}(t)-{\bf R}(t-5dt))^{2}/N)^{1/2}]$ between the current and prior IS, with no reference to the presence or absence of a transition. Distributions for $T$ = 1.10, 0.90, 0.60, and 0.40 are shown in Figure~\ref{gofd}. There is a large peak, not shown,
around $d=-6$ which identifies the numerical uncertainty of the algorithms when the IS has not changed. Transition distances exhibit a bimodal structure, with a small peak at $d \sim -4.5$. The spikes in the larger, low-$d$ lobe are not noise but represent specific, frequently occurring separations. They vanish abruptly between $T=0.60$ and $T=0.40$, the range in which we believe the crossover to hopping begins.

Not all IST are associated with diffusion. A change in state of a two-level system will change the IS. Stillinger has argued \cite{fssci} that motion among {\it megabasins} is required for diffusion, with motion among the basins comprising a megabasin non-diffusive. Transition vectors with large $d$ are most likely to reflect diffusion. Similar considerations have \cite{inm} entered the efforts to relate $D$ to $Im-\omega$ instantaneous normal modes, where {\it non-diffusive} modes must be discarded. Accordingly, we count an IST when $d$ falls in the high-$d$ lobe of $g(d)$, specifically for $d>-1.55$, marked with an arrow in Figure~\ref{gofd}. This choice has little impact on our estimate of $D$, because increasing $<\omega_{is}>$ by taking more small-$d$ IST would decrease $<({\bf \delta R}^{2})>$, leaving the product almost unchanged.

\begin{figure} 
\psfig{figure=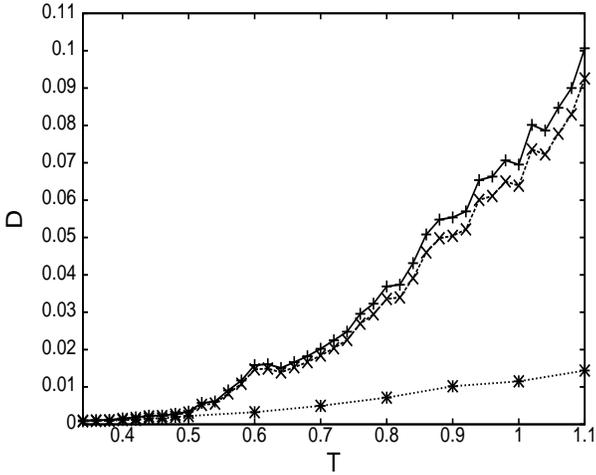,height=2.5in,width=3in,angle=-90}
\caption{Diffusion constant vs. $T$. Top to bottom at $T=1$, $D_{0}$ (IS-Markov approximation), $D_{1}$ (adjacent IST correlations), simulation. \label{ddis}}
\end{figure}

At each $T$ we obtain several IST quantities, and $D$ determined as usual from the MSD.  Figure~\ref{ddis} compares simulated $D$ with the Markov approximation $D_{0}$ and with $D_{1}$, including correlation of adjacent IST vectors. At $T \sim 1$ $D_{0} >> D$, direct evidence of
large negative correlation. However adjacent correlations only produce a $\sim 10\%$ correction, so $C(\beta)$ must be 'long ranged'.

\begin{figure} 
\psfig{figure=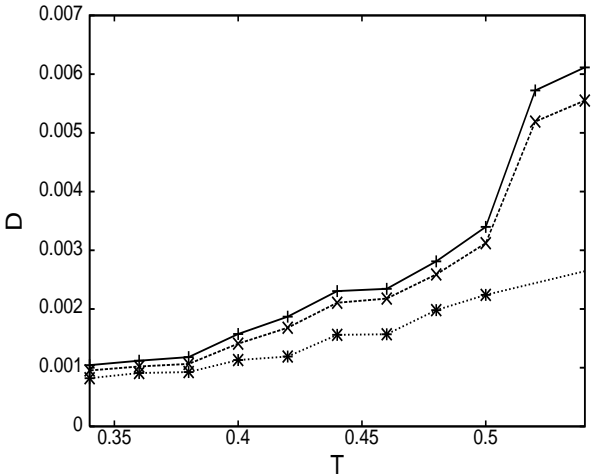,height=2.5in,width=3in,angle=-90}
\caption{Low-$T$ region of Figure~\ref{ddis}. IS-Markov approximation improves with decreasing $T$. \label{ddislo}}
\end{figure}

\begin{figure} 
\psfig{figure=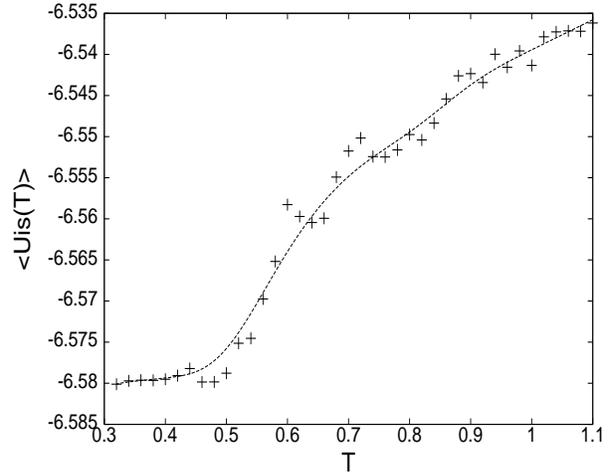,height=2.5in,width=3in,angle=-90}
\caption{Averaged IS energy vs $T$, data and Bezier fit. \label{uis}}
\end{figure}

As $T$ decreases the situation changes strikingly. While $D_{0}(T)$ initially decreases much faster than $D(T)$ and would extrapolate to zero at $T \sim 0.6$, it abruptly changes slope starting at $T \sim 0.7$ and begins to merge with $D$ at $T \sim 0.5$ (we believe the bump at $T \sim 0.6$ will vanish with more averaging). This crossover reflects behavior of $<\omega_{is}>$; $<({\bf \delta R}^{2})>$ is roughly linear for $1.10 \geq T \geq 0.34$. Figure~\ref{ddislo} highlights the region $0.54 \geq T \geq 0.34$. The joining of the curves at $T \sim 0.5$ is
clear, and $D_{0}(T)$ and $D_{1}(T)$ are approaching quantitative accuracy below $T \sim 0.4$. In the same $T$-range $<U_{is}(T)>$, Figure~\ref{uis}, is undergoing its abrupt fall to the bottom of the
landscape, previously \cite{sri} associated with crossover to hopping. Figure~\ref{uis} is similar to that given \cite{angel} by Angelani et. al. for unit-density 'modified LJ', $N=256$, where they estimate $T_{c}=0.475$.
\section{Discussion \label{disc}}
In a hopping mechanism the system is constrained by the need for activated barrier crossing to spend a long time (e.g. compared to a vibrational period) in a basin between hops; it is plausible that they should lose correlation during the wait with $D_{0} \sim D$. On the other hand at higher $T$, with thermal energy comparable to barrier heights, IST are 'bookkeeping' events as the system moves freely across IS-boundaries. As Stillinger and Weber \cite{sw} discussed, a burst of IST can then be generated by a small displacement through a region of closely spaced boundaries. If the IST rate is unpyhsically large accordingly, so too will be $D_{0}$, and negative IST-vector correlations $C(\beta)$ will be required to obtain the correct $D$.

Our results are consistent with the above scenario and suggest that the transition to a hopping mechanism occurs over the range $0.50 \geq T \geq 0.34$, possibly associated with the fall of the system to lower-lying IS. The crossover is usually identified by fitting $D(T)$ to the \cite{goetze} mode-coupling form, but the criterion $D \sim D_{0}$ is perhaps a more direct indicator. We cannot see characteristic \cite{fssci} deeply-supercooled behavior in this simulation of a simple liquid. Irregardless of any interpretation, it is very encouraging that $D_{0}$ becomes a good first approximation at the lowest available $T$. Of course correlations also exist at low $T$. If \cite{fssci,buch} the system revisits a small group of basins over and over, perhaps due to low connectivity, the corresponding IST will not cause diffusion and  $C(\beta)$ must be negative. The hope is that, for a broad spectrum of materials, such physical effects will be more tractable than the gross overestimate of D at high $T$.
\section{Acknowledgment}
We would like to thank Frank Stillinger for valuable discussions. This work was supported by NSF Grant CHE--9708055.

\end{document}